# Gold nanoparticle-pentacene memory-transistors

Christophe Novembre[1,3], David Guérin[2], Kamal Lmimouni[2], Christian Gamrat[1], and Dominique Vuillaume[2, a]

1) Commissariat à l'énergie atomique, CEA-Saclay, LIST/LCE (Advanced Computer Technologies & Architectures), Bat. 528, F-91191cedex, Gif-sur-Yvette, France.

2) Molecular Nanostructures & Devices group, Institute for Electronics, Microelectronics and Nanotechnology, CNRS & University of Lille, BP60069, avenue Poincaré, F-59652cedex, Villeneuve d'Ascq, France.

3) Commissariat à l'énergie atomique, CEA-Grenoble, MINATEC, LETI/DIHS, avenue des martyrs, F-38019cedex, Grenoble, France.

We demonstrate an organic memory-transistor device based on a pentacene-gold nanoparticles active layer. Gold (Au) nanoparticles are immobilized on the gate dielectric (silicon dioxide) of a pentacene transistor by an amino-terminated self-assembled monolayer. Under the application of writing and erasing pulses on the gate, large threshold voltage shift (22 V) and on/off drain current ratio of ~$3 \times 10^4$ are obtained. The hole field-effect mobility of the transistor is similar in the on and off states (less than a factor 2). Charge retention times up to 4500 s are observed. The memory effect is mainly attributed to the Au nanoparticles.

---

[a] E-mail : dominique.vuillaume@iemn.univ-lille1.fr

Organic devices have gained a very great interest both for the understanding of their electronic properties and for applications in low-cost, large-area, and flexible electronics. Among them, organic memories are less studied than organic field-effect transistors (OFET) and organic light-emitting diodes (OLED). Memory devices made of metal nanoparticles (NP) embedded in organic materials have been demonstrated.[1-8] These devices used a vertical structure where the active layer (an organic semiconductor including metal nanoparticles) is sandwiched between two metal electrodes. However, this approach suffers from possible drawbacks induced by the metal deposition on top of the active organic layer. For instance, it has been reported that the switching effects are due to the formation/breaking of metallic nano-filaments rather than induced by electrostatic effects due to charge/discharge of the NPs.[5, 9, 10] Here, we report on a horizontal three-terminal structure (transistor) which is not sensitive to that drawback. We demonstrate both a memory and a transistor effects combined in the same device. We report on/off ratio of $\sim 3 \times 10^4$, large threshold voltage shift (22 V) and retention time of ~4500 s.

We used a bottom-gate, bottom source-drain contacts OFET configuration. The gate is a highly-doped ($p^+$) silicon wafer covered by a thermally grown 200 nm-thick silicon dioxide ($SiO_2$). Source and drain electrodes (Ti/Au 20 nm/200 nm) were evaporated through a shadow mask. Transistor sizes are: channel length L=12μm and width W=113μm. These structures are used as is for a reference sample (sample A). For sample B, the surface was functionalized by an amino-terminated self-assembled monolayer (SAM) before the NP deposition.[11-13] First, gold (Au) electrodes were functionalized by dipping in a 2-amino ethanethiol molecules solution in ethanol (10mg/mL) during 5h. The sample was then rinsed 3 times with alcohol and subsequently dried in argon stream. Second, the SiO$_2$ surface was functionalized by immersion in a solution of (3-aminopropyl) trimethoxysilane (APTMS) molecules diluted in toluene at a concentration 1.25µL/mL and at 60°C during 4min.[14] Excess, non-reacted, molecules were removed by rinse in toluene, and then in alcohol under sonication. This sample was subsequently dried under argon stream. Static water contact angle was 19°, a common value for hydrophilic NH$_2$-terminated surfaces.[14] Sample B was then dipped in an aqueous solution of citrate-stabilized Au-NP (colloidal solution purchased from Sigma Aldrich, 20 ± 3 nm in diameter) overnight under argon atmosphere. The sample was then cleaned with deionized water and alcohol, and dried under argon stream. Finally, Au-NP were encapsulated by dipping in a solution of 1,8-octanedithiol in ethanol (10µL/mL) during 5h to help the formation of a network of NPs. The sample was subsequently rinsed in alcohol and dried in argon stream. On a second sample (sample C), Au-NPs were directly deposited onto the surface without any surface functionalization. A drop of Au-NP in aqueous solution was deposited on the substrate and the sample was left at room temperature overnight. The NP progressively deposited on the substrate as the solvent was evaporated. Sample C was cleaned in an isopropanol bath, and then in deionized water. It was subsequently dried in argon stream.

Scanning electron microscope (SEM) images, Fig. 1, show a random distribution of NPs for sample B, with the formation of some aggregates. From an image analysis, we deduce an average density of ~2-5x10$^{10}$ NP/cm$^2$. On the contrary, for sample C, we have only observed aggregates of NPs with no NP on the surface between them. An average density of 0.5-1.0x10$^{10}$ NP/cm$^2$ is roughly estimated (by dividing the area of the aggregates by the known size of the NP and multiplying this number by the average number of aggregates per cm$^2$). On the average, sample B has a higher NP density than sample C. This is expected since amino-terminated surfaces are know to react easily with Au-NP through electrostatic interactions between the negatively charged citrate overlayer of NPs and the protonated amino-terminated monolayer.[13] These three samples (including the reference, sample A) were placed together in a vacuum evaporator to deposit a 50 nm-thick pentacene film a rate of 0.1 nm/s on a substrate kept at room temperature.

Electrical measurements were done in a glove-box (nitrogen purged) using an Agilent 4155C Semiconductor Parameter Analyzer. For each type of samples, we measured about 30 devices. To investigate the memory properties, we measured $I_{DS}$-$V_{GS}$ (at $V_{DS}$= -20 V) and $I_{DS}$-$V_{DS}$ (at $V_{GS}$= -20 V) before and after writing and erasing pulses. Writing pulse consisted in applying -50 V on the gate electrode during 20s, while the source and drain electrodes were grounded. To erase the memory, we applied +50V on the gate for the same time. Figure 2 presents typical results for the devices A and B. A clear effect is demonstrated with a large reduction of the drain current and shift of the threshold voltage after the writing pulse (Figs. 2-b and c). The maximum drain current on/off ratio is 2.6x10$^4$ (at $V_{GS}$ = $V_{DS}$ = -20 V) and the maximum threshold voltage shift is $\Delta V_T$ ~ - 22 V. This on/off ratio is on a par with literature data reported for the vertical two-terminals devices.[1, 2, 4, 5] A similar effect (not shown), while lower in amplitude, is observed for device C: on/off ratio ~ 1.2x10$^2$ and $\Delta V_T$ ~ - 3.5 V (see Table I). In these two devices, these effects are larger than a very small one also observed in the reference device A (no Au-NP): on/off ~ 2.2 and $\Delta V_T$ ~ - 2 V (Fig. 2-a). For a pentacene alone OFET, these effects have been already observed and they have been attributed to defects in the pentacene films or at the gate dielectric/pentacene interface.[15] It is clear that the presence of Au-NPs induces larger effects. We can also notice that after the write operation, the slope of $(I_{DS})^{1/2}$-$V_{GS}$ characteristics does not changed too much (Fig. 2-b), meaning that the influence of writing on charge carrier mobility is small (less than a factor ~ 2). The large negative $\Delta V_T$ can be ascribed to the trapping of positive charges in the Au-NP at the pentacene/dielectric interface. With an oxide capacitance $C_{ox} = 1.7 \times 10^{-8}$ F.cm$^{-2}$, about $2 \times 10^{12}$ charges per cm$^2$ are stored in the Au-NPs, i.e. about 40-100 charges per NP for device B (calculated from the maximum $\Delta V_T$). The smaller $\Delta V_T$ for device C is consistent with the smaller NP density. An exact calculation of the charge density per NP is not very significant here due to the very crude estimate of this NP density for device C. However, we can notice that $\Delta V_T$ in device C is smaller by a factor ~ 6 than in device B, while the NP density is smaller by a factor 2-10. Given these rough estimates, these results seem consistent with almost the same order of magnitude of charges per NP in devices B and C (Table I). Moreover, some differences in $\Delta V_T$ can also come from the difference in the morphology of the NPs deposition (random network vs. large aggregates, see Fig. 1). After the erasing pulse at +50V/20s, the recovery is not complete. A complete recovery is obtained at larger bias or longer time. This asymmetry between the writing and erasing conditions can be explained by the fact that for $V_{GS} > 0$ the organic transistor is in depletion mode. Then, a space charge region appears in the semiconductor. A part of the potential drop is supported by the space charge region in the semiconductor, thus lowering the potential seen by nanoparticles at the pentacene/dielectric interface. These results show that the memory effect is due to the storage of holes in NPs. This is in agreement with charge storage properties recently demonstrated by Leong et al.[16] in a silicon/SiO$_2$/APTES/NPs/pentacene/Au vertical (capacitor-like) structure (APTES is 3-aminopropyl-triethoxysilane). In this last work, hysteresis has been observed in the capacitance-voltage characteristics, with a voltage shift of about 1-2 V achieved with writing/erasing voltages in 5-10V range. Taking into account the thinner oxide (4.5 nm in this latter case), the memory effect amplitude (~ 4-7x10$^{12}$ stored charges/cm$^2$) is on a par with our results.

We also observed that the device B (before any writing) has a more negative $V_T$ and a higher drain current than the reference device A (Figs. 2-a and b). The more negative $V_T$ may be related to the surface functionalization by the positively charged amino-terminated SAM.[17] This shift (~ - 6V compared to device A) corresponds to about 6x10$^{11}$ charges/cm$^2$. A similar effect, also observed for device C (no SAM), may be due to a surface oxide contamination by the overnight exposition to the citrate-stabilized NP solution which also contains Na$^+$ ions (which are well known as SiO$_2$ contaminants).[18]

We investigated the charge retention properties of the memory, i.e. the relaxation dynamics of trapped charge carriers after writing of the device. This relaxation was measured by connecting all the three electrodes of the device to the ground, and regularly measuring (every 20s) the drain current at $V_{DS}$ = $V_{GS}$ = - 20 V during a limited time, namely 2ms. We ensured that this measurement method induced limited charging effect, which can perturb the measurement of the charge relaxation, by checking on a non charged device. No significant charge effect has been detected for this measurement condition. Figure 3 shows the charge relaxation measured on devices B and C after writing at $V_{DS}$ = -50 V for 20 s.

We observed a slow recovery of the drain current. All the data can be fitted by the sum of two exponentials as shown by the red lines. The extracted time constants, with the dispersion limits, are given Table II. We have observed the same trends for devices C. Device A (no NP), exhibited a smaller amplitude of drain current variation and smaller time constants. We can also define a half-amplitude retention time as the time needed to reach a drain current $I_D(1/2) = I_D(0) + \frac{1}{2}[I_D(\infty) - I_D(0)]$ where $I_D(0)$ is the off current (at t=0 after the writing pulse) and $I_D(\infty)$ is the on current (at a complete recovery). These values are also summarized in Table II. Our best result (~ 4500 s, see Table II) is shorter than reported data for a vertical two-terminal memory device (~ $10^4$-$10^5$ s).[1,6] We can notice that time constants and half-amplitude retention times are higher in device B than in device C. This is consistent with the fact that device B presents a higher NPs density, and as a consequence, more charge retention centers than device C. Note that time constants for device A are of the same order of magnitude than those reported by Gu et al. for a pentacene-based OFET.[15] The physical origins of these two time constants remain unclear at the moment and deserve further studies.

In summary, we have demonstrated an organic memory combined with a transistor. The charge trapping and charge retention are attributed to the presence of a layer of Au-NPs at the pentacene/dielectric interface. A larger memory effect is observed when NPs are immobilized on the $SiO_2$ gate dielectric through an amino-terminated self-assembled monolayer, compared to NPs deposited on a naked $SiO_2$ surface. This "two-in-one" organic memory-OFET would be advantageously used in circuit design and integration for low-cost organic electronics. This may simplify memory-cell design since a charge retention capacitance is not required. Similarly, such a single device may be used to drive an OLED instead of 2 to 4 OFET. Further works are in progress to study the writing/erasing times, the cycling stability, and to address the relationship between the memory performances, the size, organization and distribution of NPs.

The authors thank R. Baptist (CEA) for his support. This work was supported by the Micro and Nanotechnology Program from French Ministry of Research under the grant RTB: "Post CMOS moléculaire".

**Table I. Threshold voltage shift, density of stored charges, stored charge per NP and on/off drain current ratios (NR stands for non-relevant). Min / max values are given, except for ΔQ and charges/NP, calculated from the maximum ΔV$_T$.**

|  | Device A | Device B | Device C | Literature results |
|---|---|---|---|---|
| ΔV$_T$ (V) | - 1.2 / - 2 | -9 / -22 | -2.7 / -3.5 | NR |
| ΔQ$_{MAX}$ (cm-2) | NR | $2 \times 10^{12}$ | $3 \times 10^{11}$ | NR |
| Charge/NP | NR | ~40-100 | ~30-60 | NR |
| On/off | 1.9 / 2.2 | 180 / $2.6 \times 10^4$ | 6 / 120 | Typical: $10^4$ (Refs. 1,2,4) |
|  |  |  |  | Best : $10^9$ (Ref. 5) |

**Table II. Time constants and retention times measured on devices A, B and C. Min / max values are given (NR stands for non-relevant).**

|  | Device A | Device B | Device C | Literature results |
|---|---|---|---|---|
| Time constant $\tau_1$ (s) | 20 / 30 | 420 / 450 | 100 / 200 | NR |
| Time constant $\tau_2$ (s) | 350 / 400 | 1500 / 3000 | 1500 / 2500 | NR |
| Half-amplitude retention time $t_{1/2}$ (s) | 140 / 270 | 1600 / 4530 | 635 / 1270 | $10^4 - 10^5$ (Refs. 1,6) |

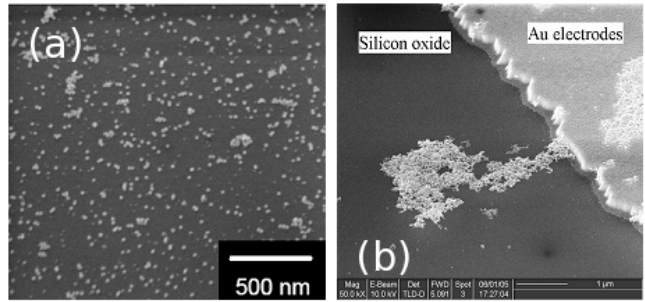

**Fig. 1 : (a) SEM images of sample B after deposition and encapsulation of Au-NP and (b) of sample C after deposition of nanoparticules. Deposition of nanoparticles on the extent of respective surfaces is similar to the presented images. On both images light areas represent 20nm Au nanoparticles, and the darker background is the oxide substrate.**

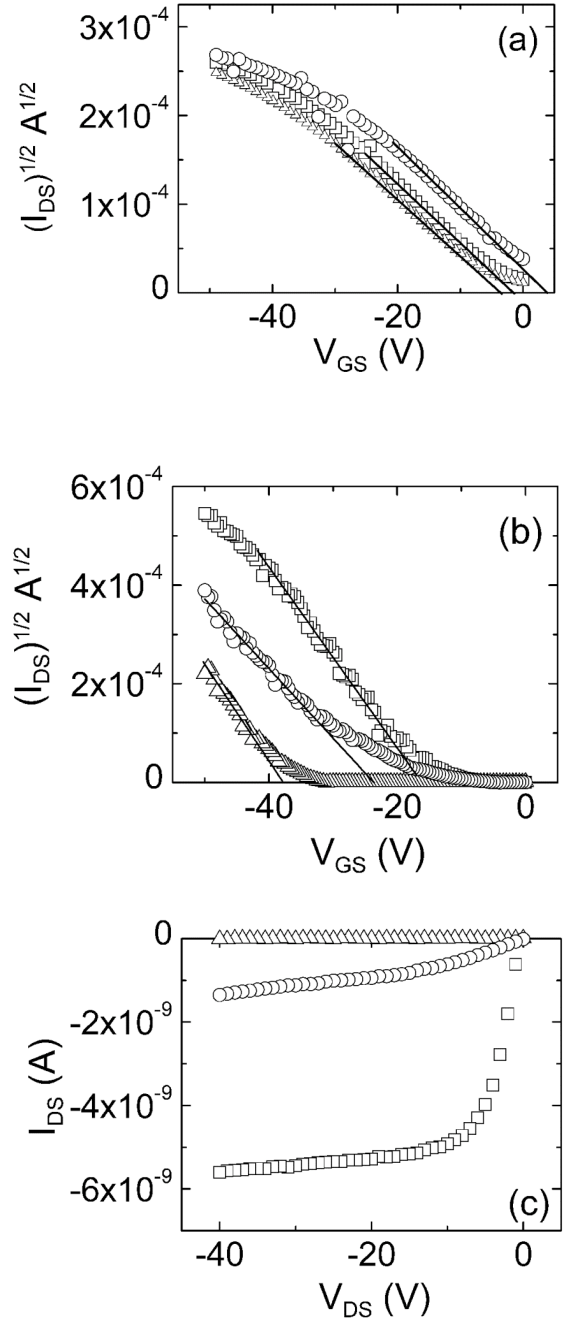

FIG. 2. $(I_{DS})^{1/2}$-$V_{GS}$ (at $V_{DS}$ = -20V) for (a) device A and (b) device B and (c) $I_{DS}$-$V_{DS}$ (at $V_{GS}$ = -20V) characteristics for device B before (□) and after (△) writing (at $V_{GS}$ = -50V, 20s) and (O) erasing (at $V_{GS}$ = 50V, 20s) operations. Solid lines on the $(I_{DS})^{1/2}$-$V_{GS}$ curves are linear extrapolations to determine the threshold voltage $V_T$.

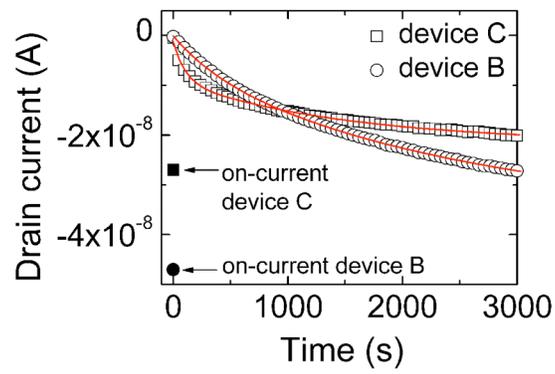

**FIG. 3. Charge relaxations in devices B and C after writing at $V_{GS}$ = -50V, 20s. Open symbols are the initial on-current before writing. The full lines are fits with two exponentials.**